\documentclass[12pt]{article}

\pdfoutput=1

\usepackage{geometry}
\usepackage{graphicx}
\usepackage{amsfonts,amssymb,amsmath,amsthm}
\usepackage{color}
\usepackage{authblk}

\definecolor{libl}{cmyk}{0.2,0.1,0,0}

\newtheorem{Theorem}{Theorem}

\newtheorem{Def}{Definition}

\newcommand{\ncd}{\newcommand}
\ncd{\lce}{\equiv_{l.C.}}
\ncd{\cg}{\mbox{cg}}
\ncd{\Tr}{\mbox{Tr}}
\ncd{\CNOT}{\mbox{CNOT}}
\ncd{\cPhase}{\mbox{cPhase}}
\ncd{\CS}{|\phi\rangle_{{\cal{L}}_2}}

\def\ket#1{\vert#1\rangle}

\def\ipr#1#2{\langle#1\vert#2\rangle}

\def\Longarrow{\protect\@lra}
\def\@lra{\relbar\joinrel\relbar\joinrel\relbar\joinrel%
          \relbar\joinrel\rightarrow}

\newcommand{\bc}{\begin{center}}
\newcommand{\ec}{\end{center}}
\newcommand{\be}{\begin{equation}}
\newcommand{\ee}{\end{equation}}
\newcommand{\bea}{\begin{eqnarray}}
\newcommand{\eea}{\end{eqnarray}}

\ncd{\QCcns}{$QC_{\cal{C}}$}
\ncd{\QCc}{$QC_{\cal{C}}\;$}

\begin{document}

\title{Quantum computation by local measurement}
\author[1]{Robert Raussendorf}
\author[1,2]{Tzu-Chieh Wei}
 \affil[1]{\small Department of Physics and Astronomy, University of
British Columbia, Vancouver, BC V6T 1Z1, Canada}
 \affil[2]{\small C. N. Yang Institute for Theoretical
Physics, State University of New York at Stony Brook, Stony Brook,
NY 11794-3840, USA}

\date{\today}

\maketitle

\begin{center}
  \parbox{0.9\textwidth}{\small{
 \textbf{Abstract:} Quantum computation is a novel way of information processing which allows,
  for certain classes of problems, exponential speedups over classical computation. Various models
  of quantum computation exist, such as the adiabatic, circuit and measurement-based models.
  They have been proven equivalent in their computational power, but operate very differently.
   As such, they may be suitable for realization in different physical systems, and also offer
   different perspectives on open questions such as the precise origin of the quantum speedup.
   Here, we give an introduction to the one-way quantum computer, a scheme of measurement-based quantum
   computation. In this model, the computation is driven by local measurements on a carefully chosen,
   highly entangled state. We discuss various aspects of this computational scheme, such as the role
 of entanglement and quantum correlations. We also give examples for ground states of simple Hamiltonians which enable
  universal quantum computation by local measurements.}}
\end{center}

\newpage

\section{Introduction}

Quantum computation is a promising approach to harness the laws of
quantum mechanics for solving computational problems. A particular
striking example is Shor's efficient quantum algorithm for factoring
large numbers \cite{Shor}, which breaks the RSA crypto system. After
this splendid start, the growing field has encountered numerous
challenges, some of which it has mastered, some of which it still
faces. As an example, decoherence was initially conceived as an
insurmountable obstacle to scalable quantum computation
\cite{Unruh}. However, the theory of quantum-error correction
\cite{Shor2}-\cite{AGP} and, alternatively, the scheme of
topological quantum computation \cite{Topol1,Topol2}, show that it
can in principle be overcome. Also, impressive experimental progress
has been made in recent years towards realizing quantum computers in
the laboratory \cite{Fact15}-\cite{Adi}. Yet, building a large-scale
device in the foreseeable future remains a great challenge
\cite{roadmap}.

At a fundamental level, we may ask ``Which quantum mechanical
property is responsible for the quantum speedup?'' In spite of a
number of candidates that have been proposed---such as entanglement,
superposition and interference, and largeness of Hilbert space---we
have no rigorous and generally applicable answer to this question
yet. Making progress in this direction may, in addition to deepening
our understanding of quantum computation, also lay the foundation
for the design of novel quantum algorithms. In 1948, introducing the
path integral formalism to quantum mechanics \cite{Feyn}, Richard
Feynman wrote: ``{\em{One feels like Cavalieri must have felt
calculating the volume of a pyramid before the invention of
calculus.}}'' Addressing the above questions in the theory of
quantum computation feels like that, too.

The paradigm of measurement-based quantum computation (MBQC), with
the teleportation-based schemes~\cite{GoChua} and the one-way
quantum computer~\cite{oneway,oneway2,oneway3} as the most prominent
examples, offers a new framework within which both theoretical and
experimental challenges of quantum computation can be addressed.
This article focuses on the one-way quantum computer, in which the
measurements driving the computation are strictly local. We will
discuss its prospects for experimental realization, and examine the
roles that entanglement and quantum correlations play for it.

In the one-way MBQC, the process of computation is driven solely by
local measurements, applied to a highly entangled resource state.
This is in stark contrast to the (standard) circuit model, where the
quantum is driven by elementary steps of unitary evolution,
so-called quantum gates. In the MBQC, after a highly entangled
resource state such as a 2D cluster state~\cite{RB01a} has been
created, the local systems, say qubits, are measured individually in
certain bases and a prescribed temporal order. The choice of
measurement bases specifies which quantum algorithm is being
implemented. The measurement outcomes cannot be chosen; they are
individually random. This randomness can be prevented from creeping
into the logical processing by adjusting measurement bases according
to previously obtained measurement outcomes. Finally, the
computational output is produced by correlations of measurement
outcomes.

The remainder of this article is dedicated to a few questions that
arise at this point. (i) Is MBQC experimentally feasible? -  We
discuss pro's and con's for the experimental realization of MBQC in
Section~\ref{Exp}. (ii) Why does MBQC work at all? - We provide an
explanation of the inner workings of MBQC in Section~\ref{UniPro}.
(iii) Do resource states for universal MBQC arise naturally in
quantum systems? - In Section~\ref{Cluster}, we describe how a
particular universal resource state, the cluster state, can be
realized via unitary evolution under an Ising Hamiltonian. In
Section~\ref{sec:1qubit}, we explain teleportation-based
implementations of quantum gates. Building on that, in
Section~\ref{sec:MBQCuniversal}, we explain how a CNOT gate and
general one-qubit rotations can be realized using cluster states,
leading to universality of MBQC. Section~\ref{GS} describes how
computational resources can arise as ground states of relatively
simple Hamiltonians. A resource for universal MBQC is the
Affleck-Kennedy-Lieb-Tasaki state on the honeycomb lattice; see
Section~\ref{AKLT2d}. (iv) Which role does entanglement play for
MBQC? - In MBQC, the result of the computation is obtained at the
price of consuming all or most of the entanglement initially present
in the resource state. Therefore, entanglement appears as a key
resource for MBQC. In Section~\ref{ent}, this intuition is
(partially) corroborated, and in Section~\ref{TETU} its limits are
shown. (v) Which role do quantum correlations play for MBQC? The
computational power of MBQC hinges on strong correlations among the
random measurement outcomes. These classical correlations derive
from quantum correlations in the resource state. In
Section~\ref{QMF}, we illustrate this in a specific example, by
turning the Greenberger-Horne-Zeilinger proof of Bell's theorem into
a measurement-based quantum computation.

\begin{figure}
  \begin{center}
      \includegraphics[width=7cm]{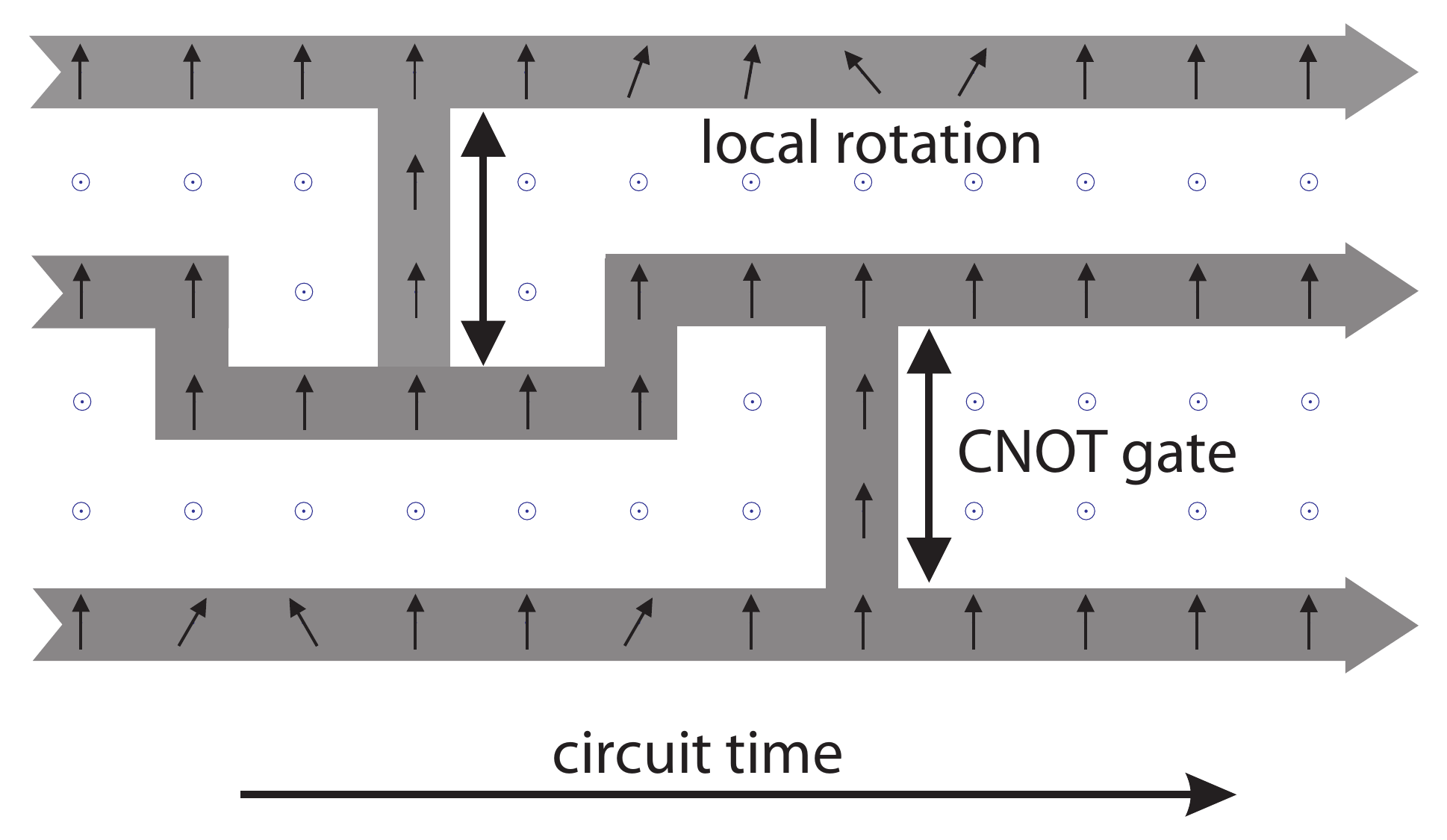}
    \caption{\label{flow} Quantum computation by measuring individual
    qubits initially prepared in a cluster state on a two-dimensional lattice,
    circuit simulator view. The choice of measurement bases specifies the sequence of simulated
    quantum gates. Circles symbolize measurements of $\sigma_z$, and arrows refer to measurement
    bases in the x-y plane.}
  \end{center}
\end{figure}

\section{Prospects for realization of scalable MBQC}
\label{Exp}

Physical systems considered for realizing a quantum computer have to
meet a set of requirements, known as the DiVincenzo criteria.
Specifically, one requires (i) A scalable setup with well-defined
qubits, (ii) The ability to initialize the qubits in a fiducial
state, say $|00..0\rangle$, (iii) A universal set of quantum gates,
(iv) The ability to measure individual qubits, and (v) Long
coherence times. For various potential realizations of a quantum
computer, such as trapped ions \cite{ITsum}-\cite{ITCM}, lattices of
cold atoms \cite{CCC,OLW}, photons~\cite{Walther} and
superconducting qubits~\cite{Nakamura}, at least a subset of the
DiVincenzo criteria have been proven in the experiment. It can be
expected that these physical systems will mature into medium-scale
test beds for quantum computers over the next couple of years, but
it is far from certain that which one of them will emerge as the
quantum counterpart of the silicon chip.

The scheme of measurement-based quantum computation can simplify the
architecture of a quantum computer since it reduces the requirements
on the interaction between qubits. First, instead of tunable
interactions between selected pairs of qubits, MBQC only requires a
translation-invariant, nearest-neighbor Ising coupling. This
interaction is highly scalable and parallelized, and requires no
control other than an on/off switch.

Physical systems that can naturally make use of this advantage are
optical lattices filled with cold atoms. In recent years,
substantial experimental progress has been made in trapping, cooling
and manipulating cold atoms in optical lattices. Large areas in such
a lattice can be regularly filled with one atomic qubit per site, by
driving a superfluid to Mott phase transition~\cite{SuMo}.
Furthermore, the Ising interaction can be realized by cold
controlled collisions between the atoms~\cite{CCC}. Finally, the
extremely difficult single site readout  has recently been
experimentally demonstrated~\cite{SiRea}.

\begin{figure}
  \begin{center}
    \includegraphics[width=10cm]{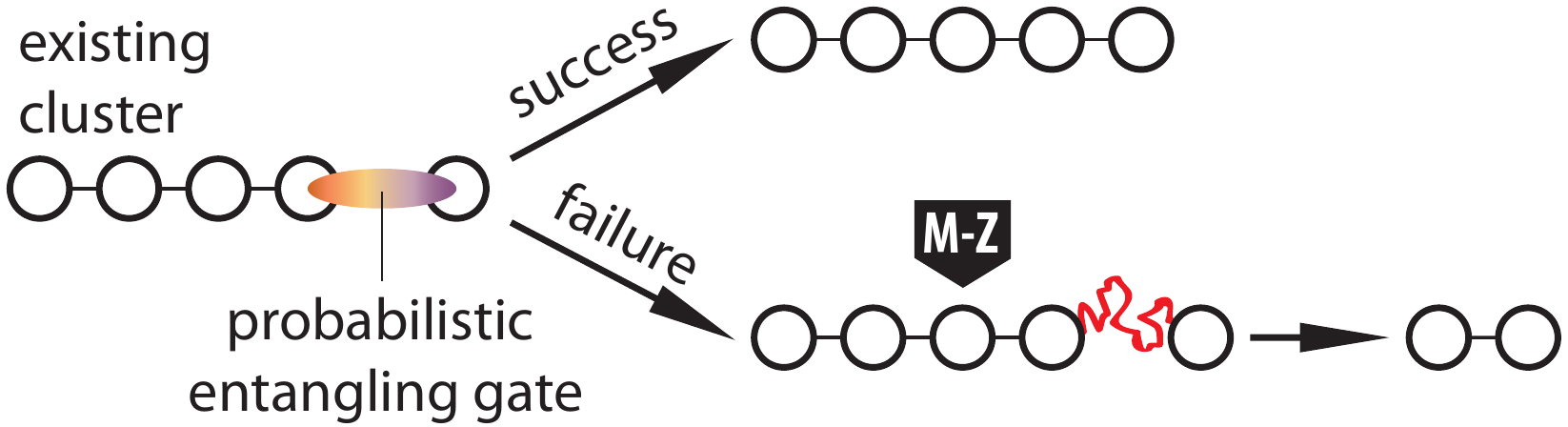}
  \end{center}
  \caption{\label{PC} Simple method for probabilistic growth of cluster states. A probabilistic heralded conditional phase gate ($\mbox{cPhase}_{ij}:=\exp(i\pi |11\rangle_{ij}\langle 11|)$) is applied to an existing linear cluster state. If the gate succeeds, the length of the cluster is increased by one. If the gate fails, the cluster qubit next to the qubits affected by the failed gate is measured in the $\sigma_z$-eigenbasis, recovering the cluster to the left. The length of the cluster is reduced by two. On average, the cluster grows if the success probability of the entangling gate is $>2/3$.}
\end{figure}

A second setting in which MBQC helps to overcome a limitation of the
interaction are probabilistic heralded entangling gates. In this
setting, the entangling gate sometimes (or mostly) fails but success
is confirmed by a classical signal. The problem with using
probabilistic gates in quantum circuits in the same way as
deterministic ones is that a single failed gate ruins the entire
computation. Instead, probabilistic heralded entangling gates may be
used to grow a cluster state. A simple protocol for probabilistic
growth of a linear cluster state  is depicted in Fig.~\ref{PC}. It
works whenever the success probability of the heralded entangling
gate is greater than 2/3. This protocol can be refined. It turns out
that cluster states of arbitrary size and geometry can be grown
efficiently for any success probability $p>0$ of the entangling
gate, enabling universal computation~\cite{Kok, DR}.

A physical setting where probabilistic heralded entangling gates
arise naturally is the Knill-Laflamme-Milburn (KLM) scheme of linear
optics quantum computation~\cite{KLM}. Probabilistic growth of
cluster states can be gainfully applied in this setting.
Specifically, it reduces the operational overhead from a factor that
grows with the size of the computation to a constant
factor~\cite{NiePrCl,BrPrCl}. See~\cite{YoRez} for a similar result
within teleportation-based quantum computation~\cite{GoChua}.
Readers interested in knowing more detail about how MBQC helps to
reduce the resource in KLM scheme beyond the explanation in
Fig.~\ref{PC} should refer to the above cited works.

MBQC also has a disadvantage for the realization of quantum
computation. As a glance at Fig.~\ref{flow} reveals, the number of
qubits that need to be stored simultaneously is significantly
increased as compared to the circuit model.  Note, however, that the
cluster state can be continuously created `on the fly', mitigating
this effect.

Finally, it shall be noted that MBQC has been realized on a small
scale in an experiment with photons~\cite{Walther}. There, a
four-qubit cluster state was created and subsequently measured,
allowing for the realization of Grover search~\cite{Grover} on a
four-item data base.

\section{How MBQC works}
\label{UniPro}

In this section we show that a  measurement-based quantum computer
has the full power of quantum computation, by mapping to the
(standard) circuit model.

To establish this result, we regard MBQC
as a circuit simulator. In this view, the cluster state provides a
`canvas' upon which a quantum circuit is imprinted by local
measurements. One spatial direction of the cluster, the vertical
direction, say, labels the positions of logical qubits on a line.
The perpendicular direction corresponds to the circuit time. As the
measurements progress from left to right, the logical qubits are
propagated  across the cluster slice by slice, implementing quantum
gates in the passing.

We assume familiarity with elementary notions of the circuit model,
such as the quantum register, quantum gates and quantum circuits
\cite{NC}. In short, a quantum circuit begins with the
initialization of an $n$-qubit quantum register in a fixed state
such as $|00...0\rangle$. Second, a sequence of unitary quantum
gates is applied. Third, the qubits of the quantum register are
individually measured in a fixed basis, and the readout of the
computation is thereby obtained.

\subsection{Cluster states}
\label{Cluster}

Of central importance for MBQC is the choice of the initial resource
state. For sure, this state has to be entangled, but not every
entangled state will do. In fact, MBQC resource states that allow
for universal quantum computation by local measurement are extremely
rare \cite{GrossFlammiaEisert,BremnerMoraWinter}, and only a small
number of such states are known explicitly.

The first universal resource state to be discovered was the
two-dimensional cluster state, which we introduce here by a
practical two-step procedure to create it. Consider a
two-dimensional lattice ${\cal{L}}_2$, with one qubit located at
each site $a \in V({\cal{L}}_2)$, the set of vertices, and with an
edge set $E({\cal{L}}_2)$, where an $e$ in $E({\cal L}_2)$ denotes a
pair of distinct vertices, e.g., $(i,j)$, that are interacting with
each other. Then, a 2D cluster state $\CS$ is created by (i)
preparing the qubits $a \in V({\cal{L}}_2)$ individually in the
state $|+\rangle_a = 1/\sqrt{2}(|0\rangle_a + |1\rangle_a)$, and
(ii) unitarily evolving this state under the Ising-like Hamiltonian
\begin{equation}
  H({\cal{L}}_2) =\hbar g\sum_{(i,j) \in E({\cal{L}}_2)}\frac{I^{(i)}-\sigma_z^{(i)}}{2}\otimes\frac{I^{(j)}-\sigma_z^{(j)}}{2}
  = \hbar g \sum_{(i,j) \in E({\cal{L}}_2)}|1\rangle_i\langle 1|\otimes |1\rangle_j\langle 1|,
\end{equation}
for a time $T=\pi/g$. Equivalently,
\begin{equation}
  \label{CScreate}
  \CS = \prod_{(a,b) \in E({\cal{L}}_2)}\mbox{cPhase}_{a,b}\, \bigotimes_{a \in V({\cal{L}}_2)} |+\rangle_a.
\end{equation}
Therein, $\mbox{cPhase}_{c,t}:=\exp\left(i\pi
|11\rangle_{c,t}\langle 11| \right)=\mbox{cPhase}_{t,c}$ is a
unitary quantum gate, a so-called conditional phase gate, a.k.a.
control-phase gate, which is symmetric w.r.t. to the control (c) and
target (t) qubits. One can also rewrite the conditional phase gate
in a useful representation
\begin{equation}
\label{eqn:cp}
 \mbox{cPhase}_{c,t}=|0\rangle_c\langle 0|\otimes I_t +
|1\rangle_c\langle1|\otimes Z_t,
\end{equation}
which only flips the phase of the target qubit when the control
qubit is in the state $|1\rangle$.
 The conditional phase
gates on the r.h.s. of Eq.~(\ref{CScreate}) commute, so that their
temporal ordering is immaterial.

The control-phase gate can be shown to satisfy the following
properties
\begin{equation}
  \begin{array}{rcl}
    \mbox{cPhase}_{a,b}\, X_a\, \mbox{cPhase}_{a,b}^\dagger &=& X_aZ_b,\\
    \mbox{cPhase}_{a,b}\, X_b\, \mbox{cPhase}_{a,b}^\dagger &=& X_bZ_a,\\
    \mbox{cPhase}_{a,b}\, Z_a\, \mbox{cPhase}_{a,b}^\dagger &=& Z_a,\\
    \mbox{cPhase}_{a,b}\, Z_b\, \mbox{cPhase}_{a,b}^\dagger &=& Z_b,
  \end{array}
\end{equation}
where, for convenience, we have used $X$, $Y$, and $Z$ to denote the
Pauli matrices $\sigma_x$, $\sigma_y$, and $\sigma_z$, respectively.
Using the above equations,  we can show that
\begin{equation}
 \prod_{ (i,j)\in E({\cal L}_2)} {\rm cPhase}_{i,j}\,X_u=\Big( X_u\prod_{v\in {\rm Nb}(u)}Z_v\Big) \prod_{(i,j)\in E({\cal L}_2)}
 {\rm cPhase}_{i,j},
\end{equation}
where ${\rm Nb}(u)$ denotes the set of vertices that are neighbors
of $u$. Applying this relation to an initial state $| +\rangle| +\rangle\cdots|
+\rangle$, we arrive at
\begin{equation}
\label{Cdef}
\Big(X_u \prod_{v\in {\rm Nb}(u)}Z_v \Big)|\phi\rangle_{{\cal
L}_2}=|\phi\rangle_{{\cal L}_2},\;\; \forall u\in V({\cal{L}}_2).
\end{equation}
Eq.~(\ref{Cdef}) uniquely specifies the cluster state $|\phi\rangle$
given the graph ${\cal L}_2$, and is thus equivalent to
Eq.~(\ref{CScreate}) as a definition for cluster states. It has the
advantage of being independent of any particular creation procedure.

Cluster states are special cases of a slightly more general class of
states, the graph states. We now define graph states in a way
similar to Eq.~(\ref{Cdef}).
\begin{Def}[Graph states and cluster states] \label{DC}Consider a graph $G$ with vertex set $V(G)$
and edge set $E(G)$, and a set of qubits, one for each vertex $a \in V(G)$. The graph
state $|G\rangle$ is the unique simultaneous eigenstate with eigenvalue 1 of the Pauli
operators
\begin{equation}
  \label{CScorr}
  K_a = X_a \bigotimes_{b \in {\rm Nb}(a) \in E(G)}Z_b,\;\;\;\; \forall a \in V(G),
\end{equation}
i.e., $|G\rangle = K_a|G\rangle$ for all $a \in V(G)$. Therein, $X \equiv \sigma_x=|0\rangle\langle 1| + |1\rangle\langle 0|$ and $Z \equiv \sigma_z=|0\rangle\langle 0| - |1\rangle\langle 1|$.
\end{Def}
A cluster state $|\phi\rangle_{\cal{L}}$ is a graph state with the
corresponding graph being a lattice ${\cal{L}}$ of some dimension
$d$, $|\phi\rangle_{\cal{L}}:= |{\cal{L}}\rangle$.

Let us now consider two examples of one-dimensional cluster states.
Example I: The two-qubit cluster state $|\phi_2\rangle$. From the
first definition Eq.~(\ref{CScreate}), using
$\mbox{cPhase}_{i,j}:=\exp(i\pi|11\rangle_{ij}\langle 11|) =
|0\rangle_i\langle 0|\otimes I^{(j)}+ |1\rangle_i\langle 1|\otimes
Z^{(j)}$, we have
\begin{equation}
\label{phi2} |\phi_2\rangle= {\rm
cPhase}_{12}|+\rangle_1|+\rangle_2=\frac{1}{\sqrt{2}} {\rm
cPhase}_{12}(|0\rangle_1+|1\rangle_1)|+\rangle_2=\frac{
|0\rangle_1|+\rangle_2+|1\rangle_1|-\rangle_2}{\sqrt{2}},
\end{equation}
where we have expanded the first qubit in $|0/1\rangle$ basis and
used the relation $Z|\pm\rangle=|\mp\rangle$ to flip the phase of
the second qubit. Although both the initial state and the
conditional phase gate are symmetric w.r.t. qubits 1 and 2, the
final state does not explicitly show this symmetry in the given basis. One may as
well use the second qubit as the control and the first qubit as the
target and redo the calculation,
\begin{equation}
\label{phi2a} |\phi_2\rangle= {\rm
cPhase}_{21}|+\rangle_1|+\rangle_2=\frac{1}{\sqrt{2}} {\rm
cPhase}_{21}|+\rangle_1(|0\rangle_2+|1\rangle_2)=\frac{|+\rangle_1|0\rangle_2+|-\rangle_1|1\rangle_2}{\sqrt{2}}.
\end{equation}
The second definition Eq.~(\ref{CScorr}) of the cluster state
$|\phi_2\rangle$ yields the relations
$X_1Z_2|\phi_2\rangle=Z_1X_2|\phi_2\rangle=|\phi_2\rangle$. We can
now explicitly verify that the state on the r.h.s. of
Eq.~(\ref{phi2}), as well as Eq.~(\ref{phi2a}), satisfies these two
relations.

Example II: The three-qubit cluster
state $|\phi_3\rangle$ on a line. Defined through Eq.~(\ref{CScreate}), it takes the form
\begin{equation}
\label{phi3}
 |\phi_3\rangle= {\rm cPhase}_{23}\, {\rm
cPhase}_{21}|+\rangle_1|+\rangle_2 |+\rangle_3={\rm
cPhase}_{23}|\phi_2\rangle|+\rangle_3=\frac{|+\rangle_1|0\rangle_2|+\rangle_3+|-\rangle_1|1\rangle_2|-\rangle_3}{\sqrt{2}}
\end{equation}
Notice that we have used the second form of
$|\phi_2\rangle$~(\ref{phi2a}), indicated by the labeling of the
conditional phase gate ${\rm cPhase}_{21}$, as then the second qubit
is in the $|0/1\rangle$ basis, convenient for the subsequent gate
${\rm cPhase}_{23}$.
 Eq.~(\ref{Cdef}) yields an implicit
but equivalent definition, through the stabilizer equations
$X_1Z_2\,|\phi_3\rangle = |\phi_3\rangle$,
$Z_1X_2Z_3\,|\phi_3\rangle = |\phi_3\rangle$,
$Z_2X_3\,|\phi_3\rangle = |\phi_3\rangle$. Again, these relations
can be explicitly verified for the state on the r.h.s. of
Eq.~(\ref{phi3}). Note that if we attach to $|\phi_3\rangle$ a
fourth qubit in $|+\rangle_4$ and apply ${\rm cPhase}_{43}$ we
obtain the linear four-qubit cluster state $|\phi_4\rangle$, and we
can continue this procedure for any linear cluster state. The number
of terms needed to describe the cluster state grows exponentially in
the number of qubits. Indeed, it doubles upon adding two more qubits
into the chain.

From the stabilizer relations follows, for example, that if qubits 1
and 3 are measured in the $Z$-basis and qubit 2 is measured in the
$X$-basis, then the measured eigenvalues $\lambda_Z^{(1)},
\lambda_X^{(2)}, \lambda_Z^{(3)} \in \{1,-1\}$ are individually
random but correlated, with $\lambda_Z^{(1)}
\lambda_X^{(2)}\lambda_Z^{(3)} =1$. In MBQC, output bits of the
computation will be inferred from correlations like this one, but
generally in more complicated bases. We note that $|\phi_3\rangle$
is locally equivalent to the so-called Greenberger-Horne-Zeilinger
(GHZ) state, $|\mbox{GHZ}\rangle = (|000\rangle +
|111\rangle)/\sqrt{2}$.  So-called one-qubit Hadamard gates have the
property that $H|0\rangle = |+\rangle$ and $H|1\rangle = |-\rangle$,
such that $H_1H_3|\mbox{GHZ}\rangle=|\phi_3\rangle$. We shall return
to the $|\phi_3\rangle$/GHZ-example in Section~\ref{QMF}, where we
establish a connection between MBQC and the GHZ-version~\cite{GHZ}
of Bell's theorem.


\subsection{Basics of quantum gates by teleportation}
\label{sec:1qubit}

In preparation for our demonstrating the universality of MBQC, we
review a few techniques of  performing gates by quantum
teleportation. We follow the discussion of \cite{CLN}. Consider a
quantum circuit
\begin{equation}
  \label{LocR1}
  \parbox{4cm}{\includegraphics[width=4cm]{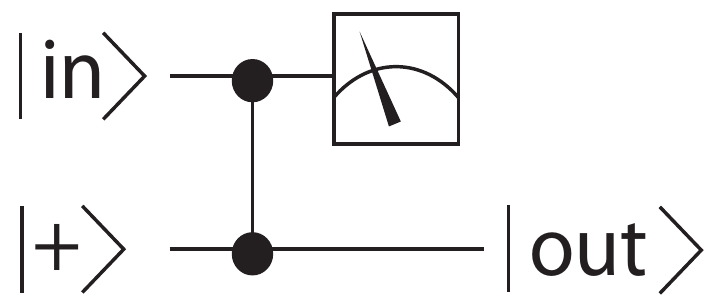}}
\end{equation}
which takes a two-qubit state $|\mbox{in}\rangle_1 \otimes
|+\rangle_2$ as input. Therein,  the first qubit is free to choose,
$|\mbox{in}\rangle = a|0\rangle + b |1\rangle$, and the second qubit
is always fixed. A conditional phase gate is applied to these qubits
and, subsequently, the first qubit is measured in the eigenbasis of
\begin{equation}
  \label{Obs1}
  O(\varphi_1) = \cos \varphi_1 \, X_1 + \sin \varphi_1 \, 2,\;\; -\frac{\pi}{2} < \varphi \leq \frac{\pi}{2}.
\end{equation}
The measured eigenvalues are $\pm 1=(-1)^{s_1}$, $s_1\in \{0,1\}$,
and the corresponding eigenstates are $|\phi_1,\pm\rangle= (|0\rangle \pm e^{i\phi_1}|1\rangle)/\sqrt{2}$. If one obtains the measurement outcome $s_1=0$, then the second qubit is projected to a
state ${}_1\langle\phi_1,+|\cdot |\psi\rangle_{12}\sim a|+\rangle_2
+ b e^{-i\phi_1} |+\rangle_2 = e^{-i\phi_1/2}(a\, e^{i\phi_1/2}
|+\rangle_2 + b\, e^{-i\phi_1/2} |-\rangle_2 )\sim H e^{i\phi_1 Z/2}
(a|0\rangle + b|1\rangle)$, up to an overall phase
$e^{-i\phi_1/2}$, where $H=(X+Z)/\sqrt{2}$ is the Hadamard
gate. In the computational basis $\{|0\rangle,|1\rangle\}$, the Hadamard gate takes the matrix form
\begin{equation}
H=\frac{1}{\sqrt{2}}\left(\begin{array}{cc}1  & 1 \cr
                          1 & -1\end{array}\right).
\end{equation}
If one obtains the measurement outcome $s_1=1$, then the second qubit is projected to a
state $\langle\phi_1,-|\cdot |\psi\rangle_{12}\sim a|+\rangle - b
e^{-i\phi_1} |+\rangle = e^{-i\phi_1/2}(a e^{i\phi_1/2} |+\rangle -
b e^{-i\phi_1/2} |-\rangle) \sim H e^{i\phi_1 Z/2} Z(a|0\rangle +
b|1\rangle)$, up to an overall phase
$e^{-i\phi_1/2}$. The two outcomes can be summarized in one equation
\begin{equation}
|{\rm out}\rangle = H e^{i\phi Z/2} Z^s |{\rm in}\rangle.
\end{equation}
The outcome of the circuit Eq.~(\ref{LocR1}) is that an `in' state
which was initially residing on qubit 1 has been re-located to qubit
2, and been acted upon by a unitary gate $H e^{i\phi Z/2} Z^s$ in
the passing.

We may now feed the state $|\mbox{out}\rangle$ into another circuit
of type Eq.~(\ref{LocR1}). The new output state $|\psi_3\rangle$,
located on a third qubit, will be related to  the initial state `in'
on the first qubit by
\begin{equation}
\label{XZr}
\begin{array}{rcl}
|\psi^{(3)}\rangle &=& H e^{i\phi_2 Z/2} Z^{s_2}
|\mbox{out}\rangle=\big(H e^{i\phi_2 Z/2} Z^{s_2}\big) \big(H
e^{i\phi_1 Z/2} Z^{s_1}\big) |\mbox{in}\rangle\\
& = & \big(Z^{s_2}X^{s_1}e^{i(-1)^{s_1}\phi_2 X/2}
e^{i\phi_1 Z/2} \big) |\mbox{in}\rangle.
\end{array}
\end{equation}
We find that in the above iterated circuit we can implement
rotations about both the $X$-and the $Z$-axis. Two more aspects are
worth of note. First, the rotation angle $\phi_x$ of the
$X$-rotation depends on the measurement outcome $s_1$ implementing
the preceding $Z$-rotation, $\phi_x=(-1)^{s_1}\phi_2$. Therefore, in
order to realize a rotation about a given angle $\phi_x$, the
measurement angle $\phi_2$---specifying the measurement basis for
qubit 2---must be adjusted according to them measurement outcome
$s_1$ obtained from qubit 1. Second, we obtain the desired rotation
$e^{i(-1)^{s_1}\phi_x X/2} e^{i\phi_1 Z/2}$ only up to a random
Pauli operator $Z^{s_2}X^{s_1}$. Such operators are called
`byproduct operators' in MBQC. Since they are known from the
measurement outcomes, in the above circuit they can be undone by
active intervention. Alternatively, they may be propagated forward
through the circuit, like $Z^{s_1}$ in the above example, flipping
rotation angles and, potentially, readout measurements in a
controlled and correctable fashion.

\begin{figure}
  \begin{center}
  \includegraphics[width=14cm]{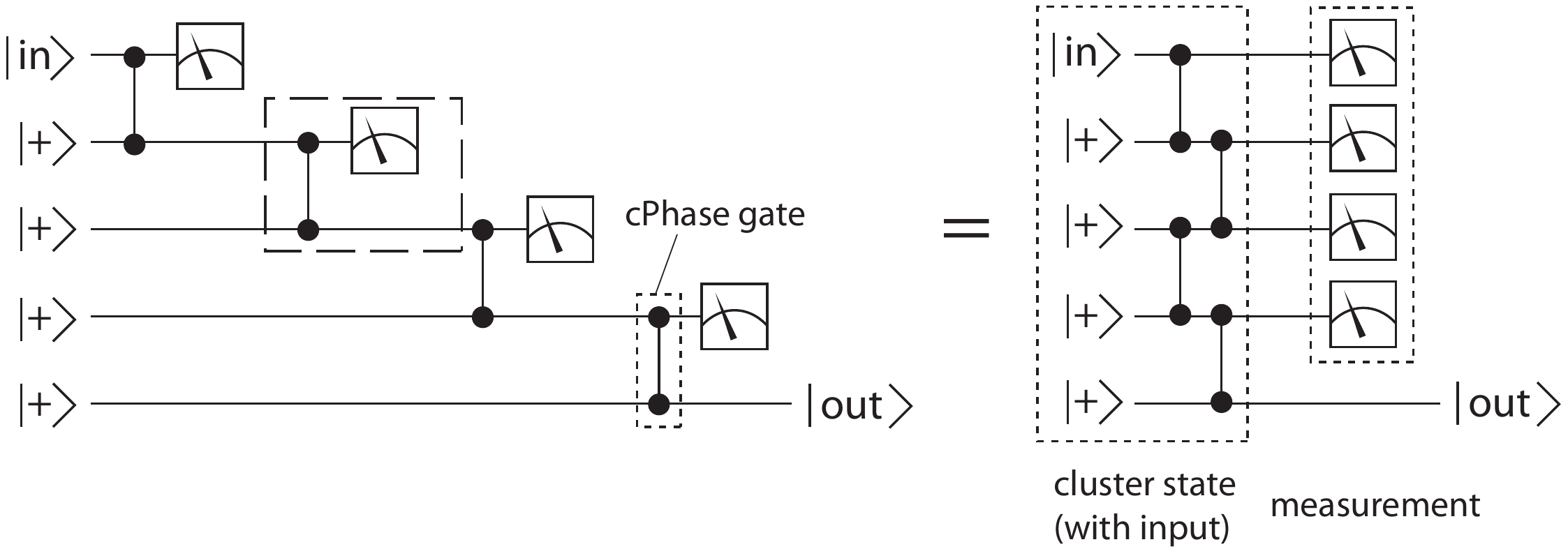}
  \caption{\label{LocRot}Circuit diagrams for MBQC-Simulation of
   a general one-qubit unitary. The horizontal lines represent the time direction of qubits. ${\rm cPhase}$ gates are
   between two qubits is indicated by the solid vertical line joining
   the corresponding two solid dots. Solid boxes denote measurements, such as the observables shown in
    Eq.~(\ref{Obs}). Lhs: the circuit Eq.~(\ref{LocR1})
   iterated four times simulates general one-qubit unitaries. Rhs:  The procedure MBQC-GS for a 5-qubt cluster state. By moving conditional phase gates past  commuting measurements, the circuits on lhs and rhs are shown to be equivalent. Thus, the procedure MQC-GS on a one-dimensional cluster of 5 qubits simulates a general one-qubit rotation.}
  \end{center}
\end{figure}

Can we build a general one qubit unitary by concatenating the
circuit of Eq.~(\ref{LocR1})?  This is indeed possible. As shown in
Fig.~\ref{LocRot}, we concatenate the circuit Eq.~(\ref{LocR1}) four
times, whereby a fifth qubit is transformed into the state
$|\psi^{(5)}\rangle=U|{\rm in}\rangle$, with the unitary gate $U$
given by~\cite{CLN}
\begin{equation}
\label{U1}
U(\{\phi,s\})=\big(H e^{i\phi_4 Z/2} Z^{s_4}\big) \big(H e^{i\phi_3
Z/2} Z^{s_3}\big) \big(H e^{i\phi_2 Z/2} Z^{s_2}\big) \big(H
e^{i\phi_1 Z/2} Z^{s_1}\big).
\end{equation}
We set $\phi_1=0$. Then, by reordering operations in the same way as
in Eq.~(\ref{XZr}), can rewrite the resulting unitary as
\begin{equation}
  \label{RotSim}
  U(\{\phi_2,\phi_3,\phi_4,s\})
  = Z^{s_1+s_3} X^{s_2+s_4} \exp\left(i(-1)^{s_1+s_3}
  \frac{\varphi_4}{2}X\right) \exp\left(i(-1)^{s_2}\frac{\varphi_3}{2}Z\right)
  \exp\left(i(-1)^{s_1}\frac{\varphi_2}{2}X\right).
\end{equation}
Therein, we have used the identities $HZ=XH$ and $XZ=-ZX$, and
dropped an overall phase factor. Thus, up to byproduct operator
$U_\Sigma \equiv Z^{s_1+s_3} X^{s_2+s_4}$, a general one-qubit
rotation
\begin{equation}
 U_{\rm rot}=
 \exp\left(-i\frac{\zeta}{2}X_i\right)\exp\left(-i\frac{\eta}{2}Z_i\right)\exp\left(-i\frac{\xi}{2}X_i\right),
\end{equation}
with Euler angles $\zeta$, $\eta$, $\xi$ can be realized by the choosing the measurement angles
\begin{equation}
  \label{infl}
  \begin{array}{rcl}
  \varphi_1 &=& 0,\\
  \varphi_2 &=& -(-1)^{s_1}\xi,\\
  \varphi_3 &=& -(-1)^{s_2}\eta,\\
  \varphi_4 &=& -(-1)^{s_1+s_3}\zeta.
  \end{array}
\end{equation}
We find that measurement angles, and thus measurement bases, depend
on measurement outcomes of other qubits. This is the origin of
temporal order in measurement-based quantum computation.

The reader will have noted that we could have accomplished the same
task of implementing a general one-qubit unitary by concatenating
the circuit Eq.~(\ref{LocR1}) only three times rather than four, and
not setting the first measurement angle to zero. However, then the
Hadamard gates in the counterpart of Eq.~(\ref{RotSim}) would not
cancel. We would still obtain a general one-qubit unitary, albeit
not in Euler normal form.


\subsection{MBQC is universal}
\label{sec:MBQCuniversal}

To prove universality of MBQC, we need to show that (i)
A universal set of gates can be simulated, (ii) Gate
simulations compose in the same way as the gates themselves, and (iii) Cluster
qubits not required in a particular computation can be removed.

\subsubsection{Simulating a universal set of gates}

We need to be able to simulate a so-called universal set of gates.
Such gate sets have the property that {\em{any}} unitary
transformation on an $n$-qubit Hilbert space, for any $n \in
\mathbb{N}$, can be arbitrarily closely approximated by gates from
the set.

A standard universal gate set consists of all one-qubit rotations
for each qubit and the controlled Not (CNOT) gate on any pair of
qubits \cite{NC}. These gates are defined as
\begin{equation}
  \label{UGS}
  \begin{array}{rcl}
  U_i &=& \exp\left(-i\frac{\zeta}{2}X_i\right)\exp\left(-i\frac{\eta}{2}Z_i\right)\exp\left(-i\frac{\xi}{2}X_i\right),\\
  \CNOT_{c,t} &= &|0\rangle_c\langle 0|\otimes I_t + |1\rangle_c\langle 1|\otimes X_t.\end{array}
\end{equation}
Therein, the subscripts $i$, $c$ (control), $t$ (target) are qubit
labels, and $\zeta$, $\eta$ and $\xi$ are the Euler angles
specifying the one-qubit rotation $U \in SU(2)$. $X$, $Z$ are Pauli
operators ($X \equiv \sigma_x,\, Y\equiv \sigma_y,\, Z\equiv
\sigma_z$). Note that in the above gate set, only the CNOT gate has
the power to entangle. It is equivalent, up to local unitaries, to
the cPhase gate introduced in Eq.~(\ref{CScreate}).

\begin{figure}
  \begin{center}
  \includegraphics[width=14cm]{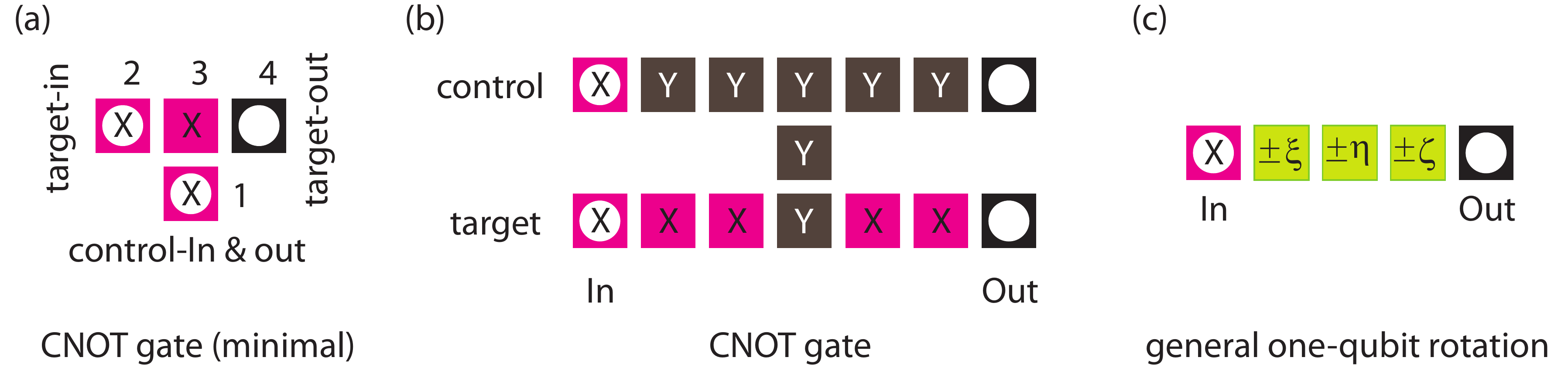}
  \caption{\label{UniGates} Measurement patterns for a universal set of gates. (a,b) CNOT,
  (c) general one-qubit rotation. The CNOT gate in (a) has the control input and output
  located on the same cluster qubit. The extended CNOT in (b) has separate locations for
  input and output qubits, for both control and target.}
  \end{center}
\end{figure}

 An MBQC can be split up into MBQC gate simulations. Each gate simulation is like a LEGO piece,
 with example patterns shown in Fig.~\ref{UniGates}, which were explained in the previous section.
 Be ${\cal{C}}$ a set of qubits, with $I \subset {\cal{C}}$ a set of
input qubits, $O \subset {\cal{C}}$ a set of output qubits, and
${\cal{C}}\backslash O$ the set of qubits which are in ${\cal C}$
but not in $O$. Then, an MBQC gate simulation on ${\cal{C}}$ is the
following \medskip

\noindent \textbf{Procedure} MBQC-GS.
\begin{enumerate}
  \item{Create a cluster state with input $|\text{in}\rangle$,
    $|\phi(\text{in})\rangle = \left(\prod_{(a,b) \in E}\mbox{cPhase}_{a,b}\right) |\text{in}\rangle_I \bigotimes_{c \in {\cal{C}}\backslash I}|+\rangle_c$.}
  \item{Measure all qubits $a \in {\cal{C}}\backslash O$, keep the state $|\text{out}\rangle_O$ of
  the unmeasured qubits in $O$. }
\end{enumerate}
The transformation $|\text{in}\rangle \longrightarrow
|\text{out}\rangle$ is  unitary if suitable local measurement bases
and sets $I,O$ are chosen. Specifically, cluster qubits which are
not measured in the eigenbasis of $Z$ are measured in a basis in the
equator of the Bloch sphere. The measured observable on such a qubit
$a \in {\cal{L}}_2$ is
\begin{equation}
  \label{Obs}
  O_a(\varphi_a) = \cos \varphi_a\, X_a + \sin \varphi_a\, Y_a,\;\; -\frac{\pi}{2} < \varphi_a \leq \frac{\pi}{2}.
\end{equation}
The angle $\varphi_a$ specifying the measured observable $O_a$ is

called the `measurement angle' for qubit $a$.

\paragraph{One-qubit rotations.} We return to the procedure of performing a general one-qubit unitary by iterating the circuit Eq.~(\ref{LocR1}) four times, c.f. Section~\ref{sec:1qubit}. By moving cPhase gates backwards in time past local measurements they commute with, we can rewrite this circuit as preparation of a cluster state with one qubit of input, followed by local measurements of four cluster qubits including the input. The necessary re-ordering is displayed in Fig.~\ref{LocRot}. The circuit on the r.h.s. of Fig.~\ref{LocRot} precisely matches the procedure MQC-GS for gate simulations, which completes the construction.

\paragraph{CNOT-gate.}

A cluster state of four qubits allows for a simulation of a
controlled-NOT gate. Consider the graph shown in
Fig.~\ref{UniGates}a. Let qubits 1 and 2 be in the states
$(a|0\rangle_1+b|1\rangle_1)$ and $(c|0\rangle_2+d|1\rangle_2)$,
respectively, and qubits 3 and 4 in the state $|+\rangle$. Now, a
control-phase gate is applied between pairs $(1,3)$, $(2,3)$, and
$(3,4)$. The joint state becomes
\begin{eqnarray}
|\psi\rangle_{1234}&=&{\rm cPhase}_{43} {\rm cPhase}_{13} {\rm
cPhase}_{23}
\Big[(a|0\rangle_1+b|1\rangle_1)(c|0\rangle_2+d|1\rangle_2)|+\rangle_3|+\rangle_4\Big], \nonumber\\
&=& \Big[a|0\rangle_1
(c|0\rangle_2|+\rangle_3+d|1\rangle_2|-\rangle_3)+ b|1\rangle_1
(c|0\rangle_2|-\rangle_3+d|1\rangle_2|+\rangle_3)\Big]|0\rangle_4
\nonumber \nonumber \\
&& +\Big[a|0\rangle_1
(c|0\rangle_2|-\rangle_3+d|1\rangle_2|+\rangle_3)+ b|1\rangle_1
(c|0\rangle_2|+\rangle_3+d|1\rangle_2|-\rangle_3)\Big]|0\rangle_4,\nonumber
\end{eqnarray}
where we have ignored overall normalization and have chosen a specific sequence of applying the commuting cPhase
gates. Let us
measure qubits 2 and 3 in the basis $|\pm\rangle$ and record the
respective eigenvalues by $(-1)^{s_2}$ and $(-1)^{s_3}$
respectively. Suppose the measurement outcomes are $+1$, $+1$, i.e.,
$s_2=s_3=0$, then the post-measurement joint state of qubits 1 and
4 is
\begin{eqnarray}
|\psi'\rangle_{14}&=&{}_{23}\langle++|\cdot |\psi\rangle_{1234}\nonumber\\
&=& a|0\rangle_1 (c|0\rangle_4 + d|1\rangle_4) + b|1\rangle_1
(c|1\rangle_4 + d|0\rangle_4)\nonumber\\
&=& {\rm CNOT}_{14}  (a|0\rangle_1+b|1\rangle_1) (c|0\rangle_4 +
d|1\rangle_4) \nonumber.
\end{eqnarray}
We see two effects: (i) there is a control-NOT
gate applied on initial qubits 1 and 2 and then (ii) the information
on qubit 2 has been transferred to qubit 4. Analyzing the three
other cases $(s_2,s_3) \in \{(0,1),(1,0),(1,1)\}$, we conclude that the output state $|\psi_{\rm
out}\rangle$ on qubits 1 and 4 is related to the
input state $|\psi_{\rm in}\rangle$ on qubits 1 and 2 via the transformation
\begin{equation}
|\psi_{\rm out}\rangle={\rm CNOT}\, X_t^{s_3} Z_t^{s_2} |\psi_{\rm
in}\rangle =Z_c^{s_2} X_t^{s_3} Z_t^{s_2}{\rm CNOT} |\psi_{\rm
in}\rangle,
\end{equation}
with qubit 1 acting as control qubit. We remark that qubits 2 and 3
are always measured in the $X$-eigenbasis, independent of all
measurement outcomes on the cluster qubits. Therefore, the
measurements on those qubits can be performed first, and the MBQC
simulation of CNOT gates entirely drops out of the temporal order of
measurements. The same holds for all gate simulations requiring only
$X$ and $Y$-measurements, such as the simulations of Hadamard gates
and rotations $\exp(i\pi/4\,Z)$.

The minimal configuration Fig.~\ref{UniGates}a for a CNOT realized
on a four qubit cluster has the property that control input and
output are located on the same cluster qubit. This may be a
disadvantage for the composition of gate simulations. The minimal
configuration can be expanded such that the locations for target
input, control input, target output and control output are all
separate; See Fig.~\ref{UniGates}b.

\paragraph{Composition of gate simulations.} It remains to be shown
that MBQC gate simulations compose like the simulated gates
themselves. This proceeds by a reordering-of-commuting-operations
argument~\cite{oneway,oneway2},\cite{CLN}; also see
Fig.~\ref{LocRot} for an example. Finally, note that the composition
of gate simulations allow for a variable input that the standard
cluster state does not provide. Starting with a cluster state
amounts to fixing the initial state of the simulated quantum
register in the state $\otimes_i |+\rangle_i$, which is the fiducial
state appearing in DiVincenzo's second criterion.

\paragraph{Removing redundant cluster qubits.}
If a two-dimensional cluster state is used to perform
measurement-based quantum computation by simulating the universal
gates as pattern of measurement (see Fig.~\ref{UniGates}), these
patterns may not cover all of the qubits on the two-dimensional
grid.  Cluster qubits such as those that are neither covered by the
gate-simulation measurement patterns  are not needed in a particular
computation can be removed by measuring them in the $Z$-eigenbasis.
Then, the remaining qubits are still in a cluster state, with the
$Z$-measured qubits removed from the cluster. This follows directly
from the creation procedure Eq.~(\ref{CScreate}) for cluster states,
and the identities $|0\rangle_a\langle 0|\,
\mbox{cPhase}_{a,b}=|0\rangle_a\langle 0| \otimes I_b$,
$|1\rangle_a\langle 1|\, \mbox{cPhase}_{a,b}=|1\rangle_a\langle 1|
\otimes Z_b$. Hence, if one measures any qubit on a cluster state in
the $Z$-basis, obtaining an outcome $0$, the remaining qubits are
exactly in a cluster state (as they are acted on by identity
operators), with the measured qubit removed from the cluster. If the
measurement outcome is $1$, then the resulting state is equivalent
to cluster state, up to local unitaries $Z$ on qubits neighboring
the measured one. This completes the proof that the MBQC can
efficiently simulate the circuit model of quantum computation.

\section{Ground states as computational resources}
\label{GS}

For $n$-qubit quantum states distributed according to the uniform
Haar measure, it has been shown that only a tiny fraction
$<\exp(-n^2)$ of states can possibly be universal resources for MBQC
\cite{GrossFlammiaEisert,BremnerMoraWinter}. We will review this
result for a different reason in Section~\ref{TETU}. Universal
resource states thus seem very rare, but is the uniform Haar measure
the right criterion to apply? Do resource states for
measurement-based quantum computation occur naturally in physical
systems, say as ground states of Hamiltonians with two-body
interactions?

Let us first drop the requirement of computational universality, and
ask the more modest question of whether ground states of suitably
simple Hamiltonians can be used as resource states for MBQC at all.
This will lead us to one-dimensional spin systems, and provide hints
for identifying a computationally universal ground state in a second
step.

\subsection{Spin chains}

In 1983, Haldane argued that the spin-$S$ Heisenberg
antiferromagnetic chain has different behaviors depending on $S$ is
integer or half-integer~\cite{Haldane}. In particular, he predicted
that when $S$ is an integer, the spin chain has a unique disordered
ground state with a finite spectral gap. This picture was supported
by a construction that Affleck, Kennedy, Lieb and Tasaki (AKLT)
proposed in a spin-1 valence-bond model~\cite{AKLT,AKLT2}. The
spin-1 AKLT model and the antiferromagnetic chain are later found to
be in the so-called Haldane phase of the following
bilinear-biquadratic model,
\begin{equation}
\label{Haldane}
H=\sum_{i} \left[ \cos\theta(\vec{S}_i\cdot
\vec{S}_{i+1})+\sin\theta (\vec{S}_i\cdot \vec{S}_{i+1})^2 \right],
\end{equation}
for $\theta \in (-\pi/4,\pi/4)$, where $\vec{S}$ denotes the spin
operators for the spin-1 particle. The one-dimensional
Affleck-Kennedy-Lieb-Tasaki (AKLT) state is a special point in the
Haldane phase with $\tan \theta = 1/3$. For periodic boundary
conditions, the ground state in the Haldane phase is unique. For a
linear chain, it is four-fold (near) degenerate, with the splitting
in the degeneracy being exponentially small in the length of the
chain. There, both edges carry an effective spin-1/2 particle. The
resulting edge states turn out to be very important for our
description of MBQC using ground states in the Haldane phase.

If the chain is terminated by a spin-1/2 particle at one end with
the additional Hamiltonian term $\sim \vec{S}\cdot
\vec{s}$, then the degeneracy is reduced to 2, resulting from an
effective spin-1/2 at the other end. The system thus carries
total spin-1/2, i.e., $S_{\rm tot}=1/2$ with the effective two
levels being $|G_0\rangle\equiv \Big| S_{\rm tot}=1/2, S_{\rm
tot}^z=1/2\Big\rangle$ and $|G_0\rangle\equiv \Big| S_{\rm tot}=1/2,
S_{\rm tot}^z=-1/2\Big\rangle$. The degenerate ground space can be
used to encode the information of a qubit:
$|\Psi\rangle=a_0|G_0\rangle + a_1|G_1\rangle$.

As a first result demonstrating the usefulness of Haldane ground
states for MBQC, it has been shown that the AKLT state allows to
simulate arbitrary single-qubit unitary gates by single-spin
measurements~\cite{Gross,Gross2,BrennenMiyake}; also
see~\cite{Chen10}. This is not sufficient for universal quantum
computation, but it provides a first connection between MBQC and
spin systems which have been studied in condensed matter physics for
completely different reasons.

Furthermore, if we slightly extend our computational model to
comprise of two primitives,  namely (i) Measurement of individual
spins (as before) and (ii) adiabatic turn-off of individual
spin-spin couplings in the Hamiltonian Eq.~(\ref{Haldane}), then a
more general result can be obtained: The usefulness of the ground
state of Eq.~(\ref{Haldane}) as computational resource extends to
the entire Haldane phase surrounding the AKLT point~\cite{Miyake}!
This is interesting in several ways. For example, away from the AKLT
point, the ground state of the spin chain is not explicitly known.
Quantum computation by local measurement on those states works
perfectly nonetheless. Furthermore, whereas quantum computation is
usually considered `high maintenance', i.e. any imperfect control of
the system Hamiltonian quickly leads the computation off track, here
we observe a feature of robustness: $\theta$ may vary largely
without affecting the functioning of the computational scheme. These
features are consequences of a symmetry-protected topological order
\cite{Symm1,Symm2} which characterizes the Haldane phase.\medskip

We now describe quantum computation by local measurements in the
Haldane phase, following the original discussion by
Miyake~\cite{Miyake}. Let us denote by $|G_0(j)\rangle$ and
$|G_1(j)\rangle$ the two degenerate ground states for a system of
spins from $j$ to $N$, and by $|\Psi(j)\rangle$ a qubit state
encoded in the ground state space, $|\Psi(j)\rangle =
a_0|G_0(j)\rangle+ a_1|G_1(j)\rangle$. Suppose we start with a chain
of $N$ spins in the state $|\Psi(j=1)\rangle$, for known $a_0$,
$a_1$. We can ask how this state is transformed when we turn off
adiabatically the coupling between the first and the second spin.
The initial value of the total spin is 1/2. After turning off the
coupling, the subsystem of spins $(2,..,N)$ by itself is effectively
a spin-1/2, residing in the ground space spanned by
$|G_0(j=2)\rangle$ and $|G_1(j=2)\rangle$. Composing it with the
spin 1 at site 1, by  angular moment addition $1/2\otimes 1=1/2
\oplus 3/2$, the final Hamiltonian has ground states with spin 1/2
and 3/2. However, while the coupling is being turned off, rotational
symmetry is maintained and the total spin conserved. The final state
is thus confined to the spin 1/2 sector. Using the Clebsch-Gordan
decomposition, one finds
\begin{equation}
|\Psi(1)\rangle \rightarrow |\Psi'(1)\rangle  = \sqrt{\frac{2}{3}} \left[\big(\frac{a_0}{\sqrt{2}}|0\rangle_{1} + a_1|-1\rangle_{1}\big) \otimes|G_0(2)\rangle -\big(a_0|+1\rangle_{1} + \frac{a_1}{\sqrt{2}}|0\rangle_{1}\big) \otimes|G_1(2)\rangle\right].
\end{equation}
After the adiabatic turning-off of the coupling, one can measure the
first spin in any orthonormal basis spanned by $|\pm 1\rangle$ and
$|0\rangle$. For example, we consider the basis
$\big\{|x\rangle\equiv(|-1\rangle -|1\rangle)/\sqrt{2},
|y\rangle\equiv(|-1\rangle +|1\rangle)/\sqrt{2},
|z\rangle\equiv|0\rangle\big\}$. After measuring spin 1 in the state
$|\beta\rangle$, the post-measurement state of the spins $(2,..,N)$
is ${}_1\langle \beta|\cdot |\Psi'(1)\rangle$. This state behaves as
if a quantum gate had acted on the system $(2,..,N)$ with the
initial state
$|\Psi(j=2)\rangle=a_0|G_0(2)\rangle+a_1|G_1(2)\rangle$. Depending
on the measurement outcome ($|\beta =|x\rangle, |y\rangle,
|z\rangle$, respectively), the resulting state is
\begin{equation}
X |\Psi(2)\rangle, \ XZ|\Psi(2)\rangle, \ Z|\Psi(2)\rangle.
\end{equation}
Therein, $X$ and $Z$ are the effective Pauli X and Z operators, with
$X=|G_0\rangle\langle G_1|+ |G_1\rangle\langle G_0|$ and
$Z=|G_0\rangle\langle G_0|- |G_1\rangle\langle G_1|$. One may
proceed to adiabatically turn off the subsequent couplings one by
one and measure the decoupled qubits in the basis $\{|x\rangle,
|y\rangle, |z\rangle\}$. Up to the above Pauli rotations, this
results in a quantum wire in which a logical qubit is propagated
forward along the spin chain.

This process is easily generalized to induce an arbitrary rotation on the
effective qubit state. For example, the basis
\begin{equation}
\label{MB}
{\cal{B}}(\alpha) = \Big\{\frac{(1\pm
e^{-i\alpha})}{2}|x\rangle_j +\frac{(1\mp
e^{-i\alpha})}{2}|y\rangle_j, |z\rangle_j\Big\}
\end{equation}
 gives rise to a
rotation about $z$-axis $R^z(\alpha)=|G_0\rangle\langle G_0| +e
^{i\alpha} |G_1\rangle\langle G_1|$, up to possible Pauli
corrections. Measurement in another basis can induce rotation about
other axes, such as $x$-axis. Therefore, arbitrary single-qubit
unitary evolution can be simulated in a chain residing in the
Haldane phase.\medskip

Let us briefly comment on the role of symmetries which protect the
Haldane phase~\cite{Symm1,Symm2} and computation on its edge states.
It has been shown that the Haldane phase is protected even in the
presence of perturbations, as long as they possess certain
symmetries such as time reversal~\cite{Symm2}, $S_k^{x,y,z}
\longrightarrow - S_k^{x,y,z}$. Now, the Hamiltonian
Eq.~(\ref{Haldane}), even with individual couplings turned off one
by one, has this symmetry. The measured observables with eigenbases
Eq.~(\ref{MB}) possess it as well, such that the Haldane phase
remains protected throughout the course of computation.\medskip

It needs to be pointed out that in the above construction, the
adiabatic switching-off of couplings is not merely a means to
extract individual spins from a `ground-state memory'. Away from the
AKLT point, turning off a coupling does real work for the
computational scheme by modifying the correlation with the edge
states. Thus, the question arises of whether the same computational
power can be obtained without the adiabatic part. I.e., are ground
states of spin chains in the Haldane phase resources for
{\em{1-qubit}} MBQC?

The answer is again affirmative. As shown by Bartlett and
collaborators~\cite{BartlettBrennen}, local measurements on a
Haldane ground state can be used to mimic a renormalization group
transformation. The resulting state is again in the Haldane phase,
with the length of the spin chain cut by a factor of three. The AKLT
state is a fixed point of this transformation to which the whole
Haldane phase is attracted. Thus, to simulate a one-qubit universal
MBQC on a ground state in the Haldane phase, a first set of local
measurements is used to bring the initial state as close as needed
to an AKLT state, and the remaining measurements simulate the
unitary gate.

\subsection{Universal MBQC with AKLT states in two dimensions}
\label{AKLT2d}

In the previous section we have identified an entire phase of ground
states which can serve as resources for restricted measurement-based
quantum computations. Beautiful and unexpected connections between
measurement-based quantum computation and condensed matter physics
have been found as a bonus. But a central question is so far
unanswered: Are there ground states of two-body Hamiltonians which
are {\em{universal}} resources for MBQC?

The answer again is `yes'. This was first established for spin 5/2
particles  on a honeycomb lattice~\cite{Chen}, with a suitably
tailored Hamiltonian. This result is important, because it had
previously been proven that cluster states, the standard resource
for universal MBQC, cannot arise as the ground state of a
Hamiltonian with only two-body interactions~\cite{Nielsen}. They can
nonetheless be closely approximated by ground states of such
Hamiltonians~\cite{RudoBart}.

What remains to be explored is whether Hamiltonians with universal
resources for MBQC as ground states can look simpler, more natural.
In this regard, it was first shown that the unique ground state of
an AKLT-like Hamiltonian for spins 3/2 on a two-dimensional lattice
Hamiltonian yields a universal resource for MBQC~\cite{QMagn}. Here,
AKLT-like means that within the two-dimensional lattice, 1D
quasi-chains are coupled via the AKLT Hamiltonian, and the coupling
between the chains is of a different type, but still two-body.

Finally, it has been shown by Miyake~\cite{HoneycombPI} and by Wei,
Affleck and Raussendorf~\cite{HoneycombUBC} that the AKLT state on
the honeycomb lattice is a universal computational resource. To
briefly review this result, let us first recall the definition of
the AKLT state on a honeycomb lattice.

The AKLT state~\cite{AKLT} on the honeycomb lattice ${\cal{L}}$ has
one spin-3/2 per site of ${\cal{L}}$. The state space of each spin 3/2 can be
viewed as the symmetric subspace of three virtual spin-1/2's, i.e., qubits. In
terms of these virtual qubits, the AKLT state on ${\cal{L}}$ is
\begin{equation}
  \label{AKLT2}
  |\Phi_{\rm AKLT}\rangle\equiv\bigotimes_{v \in V({\cal{L}})}P_{S,v}
\bigotimes_{e \in E({\cal{L}})} |\phi\rangle_e,
\end{equation}
where $V({\cal L})$ and $E({\cal L})$ to denote the set of vertices and edges
of ${\cal L}$, respectively. $P_{S,v}$ is the projection onto the symmetric
(equivalently, spin 3/2) subspace at site $v$ of ${\cal{L}}$. For an edge
$e=(v,w)$, $|\phi\rangle_{e}$ denotes a singlet state, with one spin 1/2 at
vertex $v$ and the other at $w$.

\begin{figure}
  \begin{center}
  \includegraphics[width=10cm]{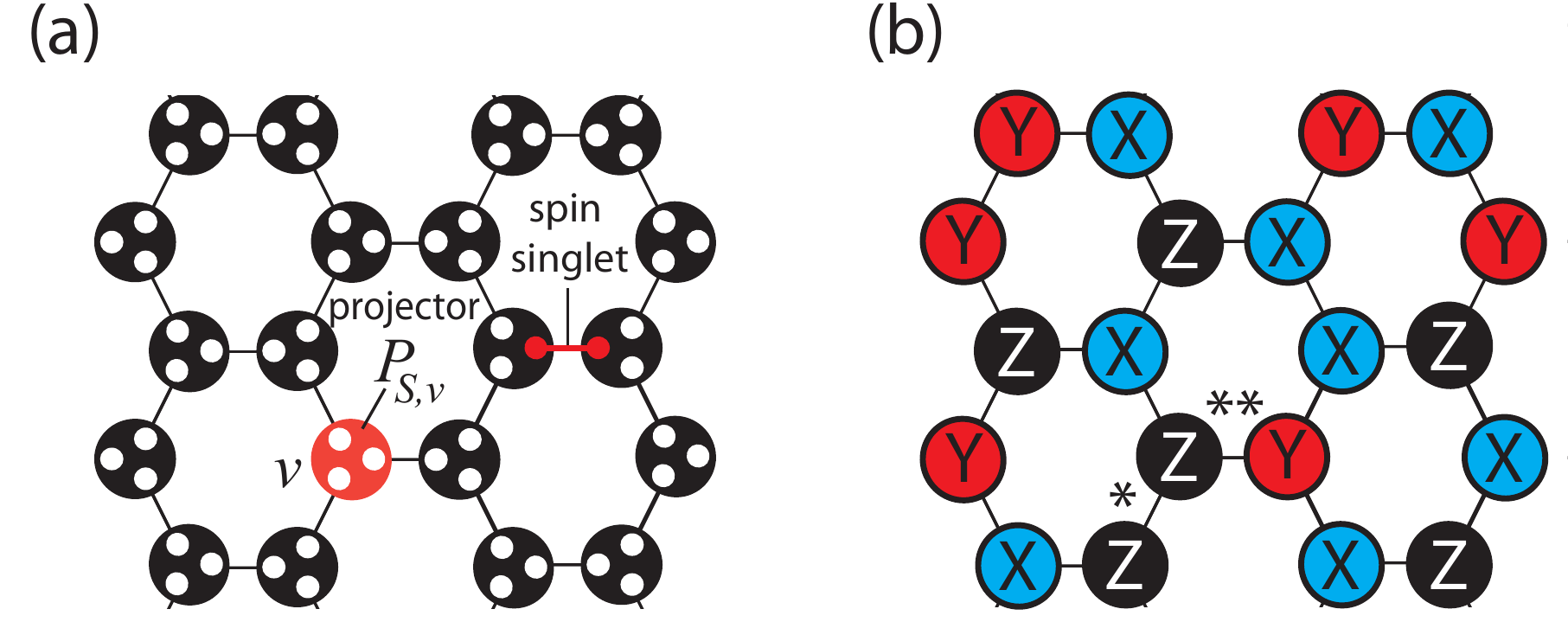}
  \caption{\label{AKLT}AKLT states. (a) Spin singlets of two virtual spins 1/2 are located on
  the edges of the honeycomb lattice. $P_{S,v}$ is a projection at each lattice site $v$ onto the
  symmetric subspace.
  (b) The first step in MBQC with AKLT states is a local POVM at each lattice site,
  with three possible outcomes  $x$, $y$, $z$. Edges with the same POVM outcome at the end points~(*)
  are treated differently from edges where the POVM outcomes differ~(**).}
  \end{center}
\end{figure}

The first step in the process of MBQC with the AKLT state is to
apply a suitable generalized measurement~\cite{NC}, also called
positive-operator-value measure (POVM), locally on every site $v$ on
the honeycomb lattice ${\cal{L}}$; See Fig.~\ref{AKLT}.
Specifically, the POVM consists of three rank-two elements
\begin{subequations}
\label{POVM2}
  \begin{eqnarray}
    {F}_{v,z}&=&\sqrt{\frac{2}{3}}
\left(\left|\frac{3}{2},\frac{3}{2},z\right\rangle \left\langle \frac{3}{2},\frac{3}{2},z\right| + \left|\frac{3}{2},-\frac{3}{2},z\right\rangle \left\langle \frac{3}{2},-\frac{3}{2},z\right|\right) = \frac{1}{\sqrt{6}}\left(S_z^2-\frac{1}{4}\right),\\
{F}_{v,x}&=&\sqrt{\frac{2}{3}}\left(\left|\frac{3}{2},\frac{3}{2},x\right\rangle \left\langle \frac{3}{2},\frac{3}{2},x\right| + \left|\frac{3}{2},-\frac{3}{2},x\right\rangle \left\langle \frac{3}{2},-\frac{3}{2},x\right|\right)= \frac{1}{\sqrt{6}}\left(S_x^2-\frac{1}{4}\right),\\
{F}_{v,y}&=&\sqrt{\frac{2}{3}}\left(\left|\frac{3}{2},\frac{3}{2},y\right\rangle \left\langle \frac{3}{2},\frac{3}{2},y\right| + \left|\frac{3}{2},-\frac{3}{2},y\right\rangle \left\langle \frac{3}{2},-\frac{3}{2},y\right|\right)= \frac{1}{\sqrt{6}}\left(S_y^2-\frac{1}{4}\right),
\end{eqnarray}
\end{subequations}
where $\textbf{r} =x,y,z$ in $|s,m_s,\textbf{r}\rangle$ specifies
the quantization axis. The above POVM elements obey the relation
$\sum_{\nu \in \{x,y,z\}}F^\dagger_{v,\nu} F_{v,\nu} = I_{S,v}$,
which is the identity on the spin-3/2 Hilbert space as well as on
the symmetric subspace of three qubits, as required. Physically,
$F_{v,a_v}$ is proportional to a projector onto the two-dimensional
subspace within the $S_a=\pm 3/2$ space, which is the origin of
logical qubits in the present construction. The outcomes $x$, $y$ or
$z$ of the POVM at the individual sites of ${\cal{L}}$ are random,
but short-range correlated. After the results of the POVM at all
sites, the post-POVM state becomes $\bigotimes_{v \in
V({\cal{L}})}F_{v,a_v}\, |\Phi_{AKLT}\rangle$, with $a_v = x,y,z$
denoting the POVM outcome at site $v$.

After using the same initial POVM, the two arguments proceed
differently. In~\cite{HoneycombPI} a mapping to quantum circuits is
pursued, whereas in~\cite{HoneycombUBC} the AKLT state is mapped to
a two-dimensional cluster state which is already known to be
universal. Let us very briefly summarize the two arguments.

In~\cite{HoneycombPI}, it is noted that whenever the POVM outcomes
on two neighboring sites differ, then the link in between can be
used to implement, along the remaining two orthogonal links, either
an elementary computational wire or an entangling gate between two
wires. Such links are sufficiently frequent to form a giant
connected  component~\cite{GT}, from which a `backbone' is chiseled
out by further local measurements. The backbone is a net composed of
computational wires and bridges between them, similar to the one
displayed in Fig.~\ref{flow}a.

The links between neighboring sites with the same POVM outcome
cannot be used for entangling gates between wires in the backbone,
and represent a (manageable) complication. A connected set of such
links ranging from one wire to another compromises those wires, and
therefore has to be avoided. Fortunately, identical POVM outcomes on
the opposite ends of a link occur only with a probability of
approximately 1/3, which is sufficiently infrequent for connected
sets of such links to remain microscopic. They can then be dealt
with by choosing a sufficiently large-meshed backbone, on which a
general quantum circuit may be implemented.

In~\cite{HoneycombUBC}, proof of computational universality proceeds
by reduction to a 2D cluster state, via an intermediate step.
Namely, it is first shown that applying the local POVMs to the AKLT
state results in a graph state. The corresponding graph depends on
the random POVM outcomes but is always planar. It is obtained from
the graph representing the honeycomb lattice ${\cal{L}}$ using the
following rules: (i) If the POVM outcomes on two neighboring
vertices agree, the edge in between is contracted, and (ii) in the
resulting multigraph, edges of even multiplicity are deleted and
edges of odd multiplicity are replaced by standard edges.

Then, typical graphs resulting from this procedure are shown by
numerical simulation to reside in the supercritical phase of
percolation, i.e., they have traversing paths. Such graph states can
then be used to implement universal quantum computation by local
measurements, which is demonstrated by reduction to 2D cluster
states.

\section{The role of entanglement}
\label{Ent}

Hardwired into the very foundations of quantum information science
is the assumption that local quantum operations and classical
communication (LOCC) are easily accomplished whereas non-local
operations, i.e., quantum-mechanical interactions between different
parts of a quantum system, are difficult. Consequently, a
fundamental distinction is made between these two classes of
operations. From this perspective,
entanglement~\cite{Entanglement,Entanglement2} is a quintessential
property of quantum systems. It measures the degree to which quantum
states require non-local operation for their creation or to which
they can enable non-local operation. It is also a key resource for
many protocols of quantum information processing, such as
teleportation, quantum cryptography, and quantum error-correction.
The defining property of entanglement monotones~\cite{VEM}, which
measure the `amount' of entanglement contained in quantum states, is
that they do not increase under LOCC.

Since MBQC is driven entirely by operations in the LOCC class,
entanglement decreases as the computation proceeds. This provides
our intuition that entanglement is a key resource for MBQC. A closer
examination shows that, for quantum computation with pure resource
states, significant entanglement is indeed necessary to achieve a
quantum speedup, but more is not necessarily better.

\subsection{Entanglement and quantum speedup}
\label{ent}

In this section, we demonstrate that any MBQC with a pure state that
only contains a small amount of entanglement can be efficiently
classically simulated, preventing a significant speedup
(See~\cite{Vidal03} for an analogous result in the circuit model).
To do so, we must first overcome an obstacle.

Consider cluster state $|\phi_{\cal{C}}\rangle$ on a one-dimensional
cluster ${\cal{C}}$. As illustrated by the example of simulating a
general one-qubit rotation, MBQC on one-dimensional cluster states
maps to the circuit model with a single qubit. It can thus be
efficiently simulated classically. And yet, $|\phi_{\cal{C}}\rangle$
is highly entangled. For suitable bi-partitions ${\cal{C}}=A \cup B$
(e.g. odd vs. even-numbered qubits), the von-Neumann entropy
$E_{A:B}(|\phi\rangle):=S(\rho_A)=-\Tr(\rho_A\log
\rho_A)=S(\rho_B)$, which is a valid entanglement measure for pure
states~\cite{Bennett}, takes the large value of
$E_{A:B}(|\phi_{\cal{C}}\rangle)=\lfloor |{\cal{C}}|/2 \rfloor$. To
rescue the asserted connection between entanglement and speedup, we
need to look for a different entanglement measure.

To this end, for general pure states $|\psi\rangle$ on a set $V$ of
qubits, consider a subcubic tree $T$ (a tree graph with vertices of
degrees between 1 and 3) whose leaves (vertices of degree 1) are
associated with the qubits in $V$. For any edge $e$ of $T$,
$T\backslash e$ consists of two components, inducing a bi-partition
of $V$ into two sets $A_T^e$ and $B_T^e$. It can be
shown~\cite{universal,universal2} that the quantity
\begin{equation}
\label{Ewidth}
E_{\rm wd}(|\psi\rangle)\equiv \min_T \max_{e\in T} E_{A_T^e,B_T^e}(|\psi\rangle).
\end{equation}
is an entanglement monotone. It is called `entanglement width'.

\begin{figure}
  \begin{center}
  \includegraphics[width=8cm]{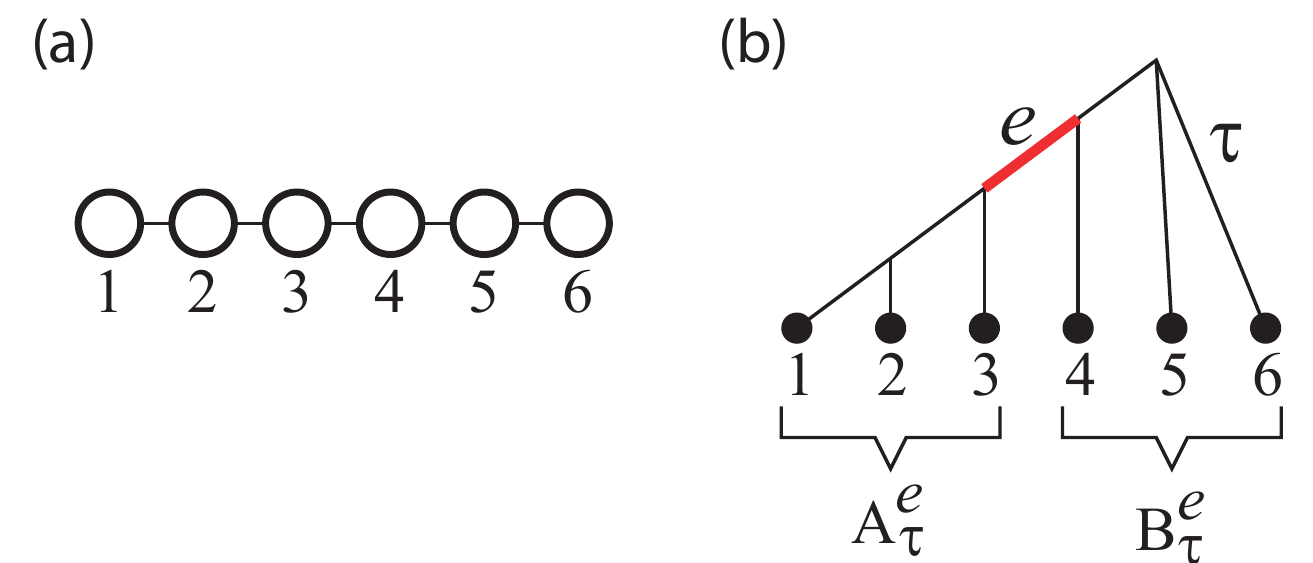}
  \caption{\label{C6}(a) Linear cluster state C6. (b) Bipartitions induced by a sub-cubic tree,
  and optimal tree $\tau$.}
  \end{center}
\end{figure}

Returning to our above example of the one-dimensional cluster state
$|\phi_{\cal{C}}\rangle$, it turns out that $E_{\rm
wd}(|\phi_{\cal{C}}\rangle)=1$. To see this, note that for the tree
$\tau$ displayed in Fig.~\ref{C6}\,b, the von Neumann entropy with
respect to the bi-partition $A_\tau^e:B_\tau^e$ is
$E_{A_\tau^e,B_\tau^e}(|\phi_{\cal{C}}\rangle) =1$, for any $e \in
E(\tau)$. Thus, the entanglement width does at least remove the
above counterexample towards establishing a connection between
entanglement and hardness of classical simulation in MBQC. But is it
of more general use? Can this entanglement measure, at least for
broad classes of interest, be efficiently calculated?

These questions both have affirmative answers. First, calculating
the entanglement width is in general hard, due to minimization over
all subcubic trees. However, if the state in question is a graph
state, then a close upper bound can be obtained
efficiently~\cite{eSpeed}, using graph theoretic
techniques~\cite{Oum}. Furthermore, the following general result
\cite{eSpeed} establishes entanglement width as the critical
complexity parameter for the classical simulation of MBQC on graph
states,
\begin{Theorem}[van den Nest {\em{et al.}}]\label{EffSim}
  Let $|G\rangle$ be a graph state on $n$ qubits. Then, MBQC on $|G\rangle$ can be classically simulated in $\mbox{poly}(n, 2^{E_{\rm wd}(|G\rangle)})$ time.
\end{Theorem}
A similar theorem can be established for general $n$-qubit quantum
states  instead of graph states only, but it requires extra
conditions relating to the efficient computability of the
entanglement width~\cite{eSpeed}.

Theorem~\ref{EffSim} shows that a substantial amount of
entanglement,  as measured by the entanglement width, is
{\em{necessary}} for a quantum speedup in MBQC with graph states.
However, it is not sufficient. MBQC with so-called surface code
states~\cite{SCS} on a $k\times k$ lattice, which are local unitary
equivalent to graph states, can be efficiently simulated classically
in $\mbox{poly}(k)$ time, but their entanglement width is linear in
$k$~\cite{BR06}.\medskip

A related yet separate question is whether substantial entanglement
is required for {\em{universality}} of MBQC. If we consider quantum
computation as a universal state preparator, so-called
$CQ$-universality, then the answer is affirmative. A family of
resource states can only be $CQ$-universal if the entanglement in
the belonging states is unbounded~\cite{universal,universal2}. In
the preceding discussion of the relation between entanglement and
speedup in MBQC, we assumed classical input and output (all qubits
were required to be measured), so-called $CC$-universality.

A $CQ$-universal quantum computer presumably has more power than a
$CC$-universal quantum computer. It was thus conjectured that
$CC$-universal resource states exist which cannot be LOCC-converted
to any $CQ$-universal state. Some examples for $CC$-universal states
were proposed by Gross and Eisert in the new framework of
measurement-based quantum computation with
Projected-Entangled-Pair-States
(PEPS)~\cite{Gross,Gross2,Verstraete}. However, it was shown later
by Cai et al.~\cite{Cai} that these $CC$-universal states can be
locally converted to cluster states, which are $CQ$-universal.
Whether or not the notions of $CC$ and $CQ$-universality are
equivalent remains an open question.

In the next section, we present a different facet of the relation
between entanglement  and MBQC which provides a counterpoint to the
result just discussed.

\subsection{Too entangled to be useful}
\label{TETU}

It came as a surprise when Gross, Flammia and
Eisert~\cite{GrossFlammiaEisert}, and independently, Bremner, Mora
and Winter~\cite{BremnerMoraWinter}, showed that if a quantum state
is {\em{too}} entangled then it becomes use{\em{less}} for MBQC.
They measured the entanglement content in terms of the geometric
entanglement (GE)~\cite{WeiGoldbart03}.
In~\cite{GrossFlammiaEisert}, it is shown that if an $n$-qubit state
$|\Psi_n\rangle$ is the resource state for a measurement-based
quantum computation that succeeds with high probability, and
furthermore $E_G(|\Psi_n\rangle)> n -\delta$, where $\delta$ is a
small constant, then this quantum computation can be efficiently
classically simulated. No quantum speedup is provided, and
$|\Psi_n\rangle$ is thus not useful as a resource state.

Let us recall the definition of geometric measure of
entanglement~\cite{WeiGoldbart03}. It is motivated by the mean-field
approximation. The idea is to find, among the set of product states,
$\{\ket{\Phi} = \ket{\phi^{[1]}}\otimes \ket{\phi^{[2]}} \otimes
\cdots \otimes \ket{\phi^{[n]}}\}$, the one closest to $\ket{\psi}$.
This is achieved by maximizing their overlap,
$\Lambda_{\max}({\psi})\equiv \max_{\Phi}|\ipr{\Phi}{\psi}|$. The GE
of $\ket{\psi}$ is then defined as~\cite{WEGM04}
\begin{equation}
E_G(\psi)\equiv -\log_2\Lambda_{\max}({\psi})^2.
\end{equation}
The maximum value of $E_G$ for an $n$-qubit state is $n$.

Now, to prove the above claim, consider MBQC on a resource state
$|\Psi\rangle$ of $n$ qubits which succeeds with high probability,
1/2 say. Furthermore, assume that $|\Psi\rangle$ has close to
maximal geometric entanglement, $E_G(|\Psi\rangle)>n-\delta$. Then,
there are $2^n$ possible measurement outcomes, and we are interested
in the fraction of good outcomes $G$ which cause the computation to
succeed. The probability of each individual outcome is bounded by
$|\langle \alpha|\Psi\rangle|^2 \le 2^{-E_G(\Psi)} < 2^{-n+\delta}$.
The success probability $\sum_{\alpha \in G}p(|\alpha\rangle)$ is
$\geq 1/2$ by assumption, and thus in turn the fraction of good
outcomes is $|G|/2^n \ge 2^{-\delta-1}$. The above quantum
computation can therefore be efficiently simulated by a classical
computer selecting the measurement outcome at random. Since the
fraction of good outcomes is large, with high probability after a
few trials, a good outcome will be selected. Thus, the above MBQC
using a (too) entangled resource state does not perform better than
a classical computer.\medskip

The conclusion that `too much entanglement renders a MBQC resource
state useless' may depend on the entanglement measure chosen. How
much so, is presently unknown. However, the GE is a lower bound on
other entanglement measures~\cite{WEGM04,HMMOV06}, such as the
relative entropy of entanglement~\cite{ER} and the logarithmic
robustness of entanglement~\cite{Robustness}. The above result holds
for those measures as well.

\section{The role of quantum correlations}
\label{QMF}

We have shown in Section~\ref{UniPro} that the computational output
in  MBQC bitwise consists of correlations among certain measurement
outcomes, and that these correlations derive from the quantum
correlations Eq.~(\ref{CScorr}) defining the cluster state. Surely,
the correlations Eq.~(\ref{CScorr}) uniquely define a highly
entangled quantum state. But is there another sense in which these
correlations reveal their quantumness? Yes.---They can in general
not be described by a local hidden variable model. Anders and
Browne~\cite{AB} have demonstrated a connection, at least in a
specific example, between measurement-based quantum computation and
Bell's theorem~\cite{Bell}.  The correlations Eq.~(\ref{CScorr})
feature prominently in this correspondence.

Hidden variable models (HVM) were spurred by Einstein,  Podolski and
Rosen's famous paper~\cite{EPR} entitled ``Can Quantum Mechanics be
considered complete?''. With no additional assumptions made, such
theories cannot be ruled out as valid descriptions of physical
reality. Bohm's wave mechanics~\cite{Bohm} is a prominent example.
However, if the hidden-variable model is required to be local,
then---as Bell's theorem~\cite{Bell} shows---it cannot reproduce all
predictions of quantum theory.

For our purpose of relating Bell's theorem to measurement-based
quantum computation, we will revisit Mermin's proof~\cite{Merm} of
Bell's theorem, using the Greenberger-Horne-Zeilinger (GHZ)
states~\cite{GHZ}. Consider the quantum state $ |GHZ\rangle =
\big(|000\rangle + |111\rangle\big)/{\sqrt{2}}$, which is local
unitary equivalent to the three-qubit cluster state
$|\phi_3\rangle$~(\ref{phi3}), as
$|GHZ\rangle=H_1H_3|\phi_3\rangle$. It is straightforward to check
that $|GHZ\rangle$ is the simultaneous eigenstate with eigenvalue +1
of the four Pauli operators
\begin{equation}
  \label{GHZcorr}
  X_1X_2X_3,\, - X_1Y_2Y_3, \, - Y_1X_2Y_3,\, - Y_1Y_2X_3.
\end{equation}
Now, a local HVM would assign `pre-existing values' $v(X_1)$, $v(X_2)$, $v(X_3)$, $v(Y_1)$, $v(Y_2)$, $v(Y_3)$ to the observables $X_1$, $X_2$, $X_3$, $Y_1$, $Y_2$, $Y_3$ which are merely revealed by measurement. Can those values be consistently assigned such that the predictions of quantum mechanics are reproduced?

Suppose this is the case. Then, $v(X_1),v(X_2), .. , v(Y_3) \in \{1,-1\}$. Furthermore,
\begin{equation}
  \label{GHZconstr}
  \begin{array}{rcr}
    v(X_1)\,v(X_2)\,v(X_3) &=& 1,\\
    v(X_1)\,v(Y_2)\,v(Y_3) &=& -1,\\
    v(Y_1)\,v(X_2)\,v(Y_3) &=& -1,\\
    v(Y_1)\,v(Y_2)\,v(X_3) &=& -1.
  \end{array}
\end{equation}
To see why these constraints need to be enforced, consider the first
one as an example. The four Pauli operators $X_1$, $X_2$, $X_3$ and
$X_1X_2X_3$ obey the identity $X_1\cdot X_2 \cdot X_3 = X_1X_2X_3$.
Furthermore, they mutually commute and hence can be simultaneously
diagonalized. The above identity therefore also holds for their
simultaneous eigenvalues, which, according to quantum mechanics, are
the possible simultaneous measurement outcomes. Since the HVM is
required to reproduce the predictions of quantum mechanics, the same
relation must hold for $v(X_1)$, $v(X_2)$, $v(X_3)$ and
$v(X_1X_2X_3)$. Finally, $v(X_1X_2X_3)=1$ by Eq.~(\ref{GHZcorr}).

But Eq.~(\ref{GHZconstr}) cannot be satisfied! Multiplying all four
equations in~(\ref{GHZconstr}) we obtain
$$
v(X_1)^2\,v(Y_1)^2\, v(X_2)^2\,v(Y_2)^2\, v(X_3)^2\,v(Y_3)^2\, = -1,
$$
Since $v(X_1)^2 = v(X_2)^2 = v(Y_3)^2 =1$, this is a contradiction.
Hence, no consistent assignment of pre-existing values $v(X_1),\, ..
\, , v(Y_3)$ exists. The correlations Eq.~(\ref{GHZcorr}) of the GHZ
state cannot be reproduced by a local HVM, and are genuinely quantum
mechanical.\medskip

As it turns out, the very same correlations power a simple yet
illuminating example of MBQC~\cite{AB}. Consider the task of
carrying out a single OR-gate via MBQC, using a GHZ-state as quantum
resource and a classical control computer for the pre-processing of
measurement bases and post-processing of measurement outcomes. This
classical processing has an important constraint, namely that---as
usual in MBQC---it can only involve addition mod 2. This kind of
computation by itself is very limited.

The computation proceeds as follows. Denote the two input bits to
the computation by $a$ and $b$, and the output bit by $o$ of the
desired OR gate, $o = a \vee b$. The observable measured on qubit
$i$ is $X_i$ if $q_i=0$, and $Y_i$ if $q_i=1$. The measurement bases
are related to the input $a,b$ via
$$
q_1 = a,\, q_2= b,\, q_3 = a + b \mod 2.
$$
The output bit $o$ is related to the measurement outcomes $s_i\in
\{0,1\}$ of the qubits $i$ via the computation by the classical
computer which can only performs AND gates (i.e., binary addition),
$$
o = s_1 + s_2 + s_3 \mod 2.
$$
Note that the relations specifying the classical pre and
post-processing are all linear mod 2, as required.

We now discuss the functioning of the MBQC-OR computer input by
input. For example, if $a=b=0$ then the measured observables are
$X_1$, $X_2$ and $X_3$. Since $X_1X_2X_3|GHZ\rangle = |GHZ\rangle$
by Eq.~(\ref{GHZcorr}), $s_1+s_2+s_3 \mod 2 =0$ for this input.
Thus, $o=0=0 \vee 0$ as required by the logical table.

As a second example, consider $a=0$ and $b=1$. Then, the measured
observables are $X_1$, $Y_2$ and $Y_3$. By Eq.~(\ref{GHZcorr}),
$X_1Y_2Y_3|GHZ\rangle = -|GHZ\rangle$, and therefore $o = s_1+ s_2 +
s_3 \mod 2 = 1=0 \vee 1$. The remaining two cases of inputs are
analogous, and the logical table of the OR-gate, $o=a\vee b$, is
established.

To put this result into perspective, it surely does not take a
quantum computer to execute an OR-gate. The present example is
therefore of no practical relevance. However, it makes a fundamental
point. The classical control computer alone, which is only able to
perform addition mod 2, has almost no computational power. In
contrast, if mod 2-addition is supplemented by the capability of
performing OR-gates, the resulting computational device becomes
classically universal. Thus, a supply of GHZ states and the ability
to measure them locally leads to a vast increase in the
computational power.

What is more, the very same quantum correlations upon which Mermin's
proof of Bell's theorem rests turn out to power the above
measurement-based quantum computation. This result~\cite{AB} hints
at a link between MBQC and non-locality of quantum mechanics. How
general this connection is remains to be explored.

\section{Conclusion}

We have given an introduction to the one-way quantum computer, a
scheme of universal quantum computation driven by local measurements
on an entangled resource state. After a short explanation of how
this scheme of computation works, we have described its underlying
computational model, and identified universal resources among ground
states of relatively simple Hamiltonians---such as the AKLT state on
the honeycomb lattice. Further, we have discussed the roles of
entanglement and quantum correlations for this computational model.
It should be noted that our knowledge in either of these areas is
very incomplete, and the research highlights presented here should
be understood as base camps for further exploration.

We would like to end with three questions of varying degree of
generality that seem particularly close to condensed matter physics:
Is there, similar to the one-dimensional case, a Haldane-like phase
around the AKLT state on the honeycomb lattice; and if so, does
computational universality extend from the AKLT state to all regions
of that phase? Can universal resource states be classified? Can a
general theory of quantum correlations for measurement-based quantum
computation be established?\medskip

 \noindent {\bf Acknowledgment}. The authors thank Maarten van den Nest,
 Dan Browne, Akimasa Miyake, Wolfgang D{\"u}r, Hans Briegel, Ian Affleck and
 Dietrich Leibfried for discussions. This work is supported by NSERC, Cifar and the Sloan Foundation.

\end{document}